\documentclass{article}
\usepackage{spconf,amsmath,graphicx,hyperref}

\usepackage{cite}
\usepackage{amsmath,amssymb,amsfonts}
\usepackage{dsfont}
\usepackage{algorithmic}
\usepackage{textcomp}
\usepackage{wrapfig}
\usepackage{kotex}
\usepackage{xcolor}
\usepackage[T1]{fontenc}
\usepackage{tabularray}
\usepackage{multirow}
\usepackage{multicol}
\usepackage{booktabs}
\usepackage{makecell}
\usepackage{tabularx}
\usepackage{subcaption}
\usepackage[table]{xcolor}
\usepackage{orcidlink}

\title{EmoFI-TTS: Face-Conditioned Expressive Voice Synthesis with Intensity Modulation}
\name{
\shortstack{
Yejin Jeon\,\orcidlink{0009-0007-8802-2675}$^{1,2,*}$, 
Luel Hagos Beyene\,\orcidlink{0009-0005-9333-4396}$^{3,4,*,\dagger}$, 
Marie Maltais\,\orcidlink{0009-0009-1362-0472}$^{1,2}$, \\ 
Hyounghun Kim\,\orcidlink{0009-0008-3382-7510}$^{5}$, 
David Ifeoluwa Adelani\,\orcidlink{0000-0002-0193-2083}$^{1,2,6}$
}
\thanks{$^{*}$Equal contribution. $^{\dagger}$Work done at Mila--Qu\'ebec AI Institute.}}

\address{
$^1$Mila-Qu\'{e}bec AI Institute
$^2$McGill University
$^3$NM-AIST \\
$^4$AIMS Research and Innovation Center 
$^5$CSE \& GSAI, POSTECH
$^6$Canada CIFAR AI Chair
}

\begin{document}

\maketitle
\begin{abstract}
Controlling emotional intensity remains challenging in expressive text-to-speech synthesis, particularly when emotion must be inferred without explicit labels. We propose EmoFI-TTS, a face-conditioned framework that derives emotion category and intensity from a single facial image while extracting speaker characteristics from reference audio. Modality-specific representations are fused to condition a pretrained speech synthesizer, with cross-speaker training encouraging separation of affect and vocal identity. An intensity-conditioned refinement module enhances the generated mel-spectrogram, while interpolation of learned intensity embeddings enables continuous control without additional training. On speaker-disjoint MEAD evaluation, EmoFI-TTS\footnote{https://github.com/luelhagos/Emofi} achieves on average 64.4\% and 75.2\% objective and subjective emotion intensity accuracies, while retaining low word error rate and high naturalness. 
\end{abstract}
\section{Introduction}
\label{sec:introduction}
Recent advances in text-to-speech (TTS) have substantially improved the naturalness and intelligibility of synthetic speech~\cite{popov2021gradtts,multispeaker2}. This progress has shifted attention toward more human-like speech generation by modeling paralinguistic factors such as emotion, with applications spanning conversational agents, dubbing, and audiobook narration. Yet achieving precise emotive control that generalizes across linguistic content and speaker identity, without sacrificing naturalness or intelligibility, remains an open challenge.

Existing expressive TTS systems typically condition emotion on manually specified labels, learned style embeddings, or emotional reference speech~\cite{shi2025emotionAligned}. Such conditioning is effective in constrained settings but ill-suited to interactive use: it demands either a hand-set emotional tag for every utterance or a recording of the target speaker already exhibiting the desired emotion. To reduce this reliance on manual specification, recent work has used large language models (LLMs) to infer a speaker's emotional state from dialogue history~\cite{jeon-etal-2025-prompt}. This strategy, however, inherits the ambiguity of text itself: identical sentences can be uttered with different, even conflicting, emotional intent. More fundamentally, none of these approaches capture how emotion actually unfolds in real communication, where a single category such as \emph{happy} spans a continuum from mild to exuberant, each level associated with markedly different prosody, energy, and spectral emphasis.

\begin{figure*}[t]
  \centering
  \includegraphics[width=0.82\textwidth]{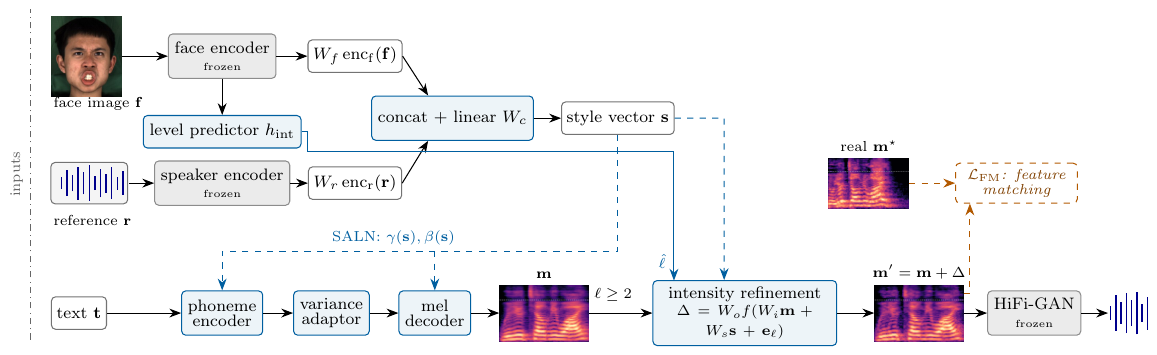}
  \caption{System overview. Text, face image, and reference audio each control content, emotion and its intensity, and speaker timbre, respectively. Solid lines carry signal, blue dashed lines carry conditioning, and the orange arrow is the feature-matching loss against the real mel $\mathbf m^{\star}$, used only in training. At inference, the emotion intensity level is predicted from the face.}
  \label{fig:arch}
\end{figure*}

Facial expression offers a natural resolution to these limitations~\cite{li2018deepface}. In face-to-face communication, a speaker's expression sets an auditory expectation before a word is even heard, priming the listener for how the utterance should sound. Unlike text or manually assigned labels, a facial expression conveys jointly \emph{which} emotion is intended and \emph{how intensely} it is felt. This observation motivates an alternative conditioning paradigm, in which a facial image supplies the affective intent of an utterance directly. 
While prior work on face-to-speech synthesis has begun to explore visual conditioning~\cite{jeon-etal-2025-progressive,ye2025emotionalfaceto}, they have predominantly treated the face as a proxy for speaker attributes such as timbre, leaving its affective role underexplored. Additionally, existing methods do not model the fine-grained emotion intensity, despite its centrality to perceived expressiveness: a mildly happy face and an exuberantly happy face are often mapped to the same high-level category, yet warrant markedly different acoustic realizations. Modeling emotion category alone can therefore yield speech that is semantically plausible but perceptually flat or mismatched to the visual cue.

To address this gap, we propose \textbf{EmoFI-TTS} (Emotion-Face-Intensity-TTS), an intensity-aware cross-modal framework for emotional speech synthesis. Given a facial image and a reference utterance, EmoFI-TTS extracts an affective category and intensity representation from the face and a timbre representation from the audio. Cross-modal fusion between the two modalities then conditions the backbone TTS model to generate an initial mel-spectrogram. An intensity-conditioned modulation module then refines the mel-spectrogram, reinforcing emotion-relevant acoustic patterns while preserving speaker consistency. This refinement stage acts as a targeted acoustic enhancement step, converting otherwise under-expressed predictions into speech whose emotional strength faithfully matches the facial cue.

This work makes three contributions. First, we formulate emotional speech synthesis as a visually-conditioned, disentangled cross-modal problem, in which facial information governs emotion category and intensity while reference audio independently governs target-speaker identity. To the best of our knowledge, this is the first work to explicitly model facial emotion intensity for face-conditioned emotional speech synthesis, moving beyond coarse categorical conditioning toward fine-grained control over expressive strength. Second, the proposed pipeline enables expressive speech generation and emotion transfer to unseen speakers without requiring emotional recordings from the target speaker. Third, subjective and objective evaluations show that this fine-grained expressivity can be controlled without sacrificing the intelligibility or naturalness of the generated speech.

\section{Method}
\label{sec:method}

\textbf{Problem Formulation.} Given text $\mathbf{t}$, a face image
$\mathbf{f}$, and a reference utterance $\mathbf{r}$, we synthesize speech $\hat{\mathbf{x}}$ whose content follows $\mathbf{t}$, emotion
category and intensity follow $\mathbf{f}$, and voice follows
$\mathbf{r}$. The backbone is a pretrained multi-speaker FastSpeech2
model~\cite{min2021metastylespeech} $G$, which generates speech conditioned on target text and a speaker's reference audio (Fig.\ref{fig:arch}).

\noindent\textbf{Affect and Timbre Learning}
To infer emotion category and intensity from a speaker's facial expression, we train an ImageNet-pretrained ResNet-50~\cite{he2016resnet} encoder $\mathrm{enc_f}$ that maps a single face frame to a $768$-d embedding, $\mathbf{e}_f=\mathrm{enc_f}(\mathbf{f})$.
Training aligns this embedding against the audio-derived emotion representation of the corresponding training speaker using emotion2vec \cite{ma2024emotion2vec}, jointly with auxiliary
emotion- and intensity-classification heads. Once trained, encoder $\mathrm{enc_f}$ is frozen thereafter. Independently, a frozen ECAPA-TDNN~\cite{desplanques2020ecapa} with least-squares erasure \cite{belrose2023leace} $\mathrm{enc_r}$ maps the reference utterance to a $192$-d timbre embedding $\mathbf{z}=\mathrm{enc_r}(\mathbf{r})$. This separation is deliberate: the face branch is trained to carry only affective information, while the audio branch is trained to carry only speaker identity. The two embeddings are then concatenated and linearly projected into a single style vector: 
\begin{equation}
  \mathbf{s} = W_c\,\big[\,W_f\,\mathbf{e}_f \,;\; W_r\,\mathbf{z}\,\big] \in \mathbb{R}^{128},
  \label{eq:arch}
\end{equation}
which conditions the phoneme encoder and mel decoder of $G$ through affine modulation.

As such, given text and style vector, $G$ produces an emotionally salient mel-spectrogram $\mathbf{m} = G(\mathbf{t},\mathbf{s})$ in the target speaker's voice. The
model is trained with the backbone model's original mel, pitch, energy, duration, and adversarial losses, and two auxiliary terms for affective information learning:
\begin{equation}
  \mathcal{L} = \big(\mathcal{L}_{\text{mel}} + \mathcal{L}_{p} + \mathcal{L}_{e}
  + \mathcal{L}_{d} + \lambda_{\text{adv}}\mathcal{L}_{\text{adv}}\big)
  + \lambda_{\text{emo}}\mathcal{L}_{\text{emo}}
  + \lambda_{\text{int}}\mathcal{L}_{\text{int}},
  \label{eq:backbone}
\end{equation}
where $\mathcal{L}_{\text{emo}}$ and $\mathcal{L}_{\text{int}}$ are
cross-entropy losses computed by small classification heads reading $W_f\mathbf{e}_f$ only ($\lambda_{\text{emo}}{=}\lambda_{\text{int}}{=}0.1$).

To further enforce the separation between affective and timbre learning, half of training uses
\emph{cross-speaker} pairing: the input face is swapped for a \emph{different} speaker's, matched only on ground-truth emotion category and intensity label, while the reference audio continues to identify the original target speaker. Since identity information from the face is now mismatched
with the target voice, it cannot serve as a shortcut for reconstructing timbre, forcing speaker identity to route exclusively through
$\mathrm{enc_r}$.

\noindent\textbf{Intensity-Aware Refinement.}
Since the decoder is trained to predict a conditional-mean mel-spectrogram, it tends to over-smooth the spectral texture, i.e., harmonicity, spectral flux, and energy, that distinguishes weak from strong emotional expression. As a
result, emotion \emph{category} is well reproduced, but \emph{intensity} is often under-rendered. This is addressed with a refinement module $R$ that is applied after mel-spectrogram generation.
Specifically, the intensity-classification head from training is frozen and then reused as
an intensity predictor $h_{\text{int}}$ and the target level
$\ell\in\{1,2,3\}$ is inferred directly from the face embedding: 
$\hat\ell = \arg\max_\ell\, h_{\text{int}}(W_f\mathbf{e}_f)_\ell$. 
$R$ then
predicts a residual on the initial mel-spectrogram through a level-gated correction:
\begin{equation}
  \mathbf{m}' = \mathbf{m} + \mathds{1}[\ell\!\geq\!2]\,\Delta, \quad
  \Delta = W_o\, f\big(W_i\mathbf{m} + W_s\mathbf{s} + \mathbf{e}_\ell\big),
  \label{eq:postfilter}
\end{equation}
where $f$ is a stack of dilated convolutions, $\mathbf{e}_\ell$ is a learned embedding of the intensity level, and $W_o$ is zero-initialized so that $R$
starts as an identity map. Note that the level enters the nonlinearity $f$ rather than scaling $\Delta$ directly, so each level produces a qualitatively distinct correction rather than a uniform gain. Lowest intensity level
($\ell{=}1$) bypasses $R$ entirely, since applying refinement was found to measurably over-energize already-calm speech.

\begin{table}[t]
  \caption{Results across objective and subjective (\texttt{Sbj}) metrics. MOS scores are reported with 95\% confidence intervals.}
  \label{tab:combined_all}
  \centering
  \resizebox{\columnwidth}{!}{%
  \begin{tabular}{l c | c c | c c c c c c}
    \toprule
    \textbf{Model} & \textbf{WER} & \textbf{UTMOS} & \textbf{MOS} & \textbf{L1} & \textbf{L2} & \textbf{L3} & \textbf{L1-Sbj} & \textbf{L2-Sbj} & \textbf{L3-Sbj} \\
    \midrule
    Real (GT) & 6.6  & 2.86          & 3.98 $\pm$0.20         & 69.7 & 34.6 & 80.3          & 88.7          & 81.2       & 87.6 \\
    EmoSpeech & 9.3  & 2.88          & 2.96 $\pm$0.26         & 42.6 & 56.0 & 18.8          & 63.5          & 46.3          & 33.8 \\
    FaceTTS   & 16.3 & 2.46          & 2.65 $\pm$0.28         & 43.1 & 44.9 & 14.9          & 61.9          & 44.0          & 40.1 \\
    \midrule
    \rowcolor{gray!25}
    Ours      & 6.9  & 3.14 & 3.25 $\pm$0.26 & 76.4 & 42.8 & 74.1 & 82.2 & 69.5 & 73.9 \\
    \bottomrule
  \end{tabular}%
  }
\end{table}

\begin{figure}[t]
  \centering
  \includegraphics[width=0.85\columnwidth]{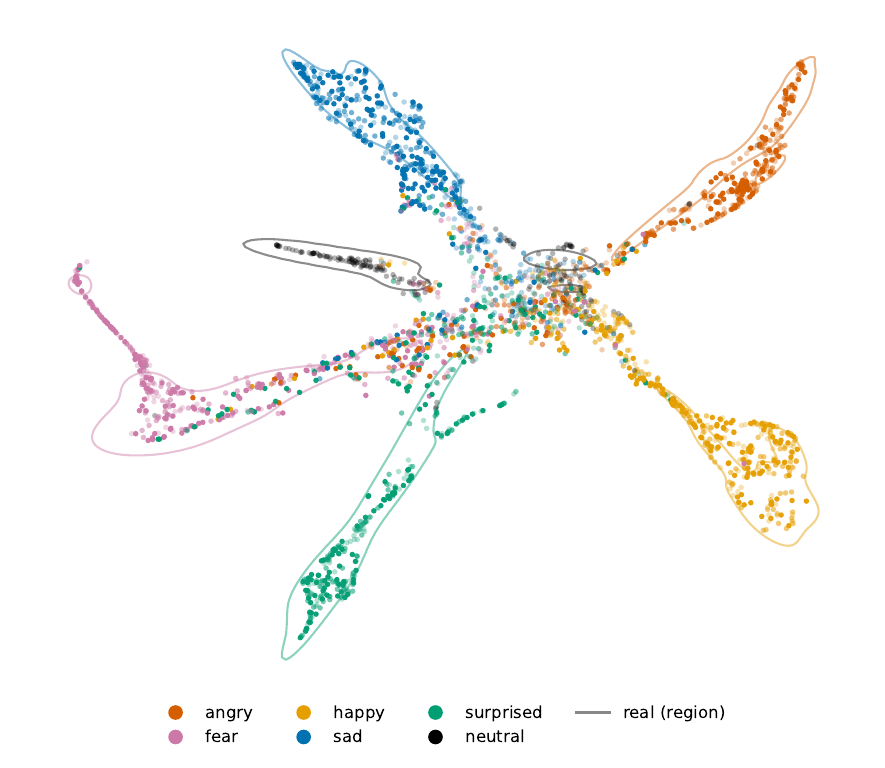}
  \caption{Emotion visualizations of generated audio.}
  \label{fig:clusters}
\end{figure}

$R$ is trained with a weighted sum of four terms: a multi-scale LSGAN loss \cite{LSGAN_loss};
a variance-matching term that restores the per-bin spectral variance; an $\ell_1$ content anchor
$\lVert\mathbf{m}'-\mathbf{m}\rVert_1$; and a speech-emotion feature-matching loss,
\begin{equation}
  \mathcal{L}_{\mathrm{FM}} = \big\lVert \phi(\mathrm{voc}(\mathbf{m}')) -
  \phi(\mathrm{voc}(\mathbf{m}^\star)) \big\rVert_1,
  \label{eq:fm}
\end{equation}
where $\phi$ is an intermediate layer of a frozen
wav2vec~2.0~\cite{baevski2020wav2vec}, $\mathbf{m}^\star$ is the ground-truth
mel-spectrogram, and both sides are passed through the same frozen universal HiFi-GAN vocoder~\cite{kong2020hifigan}.
Although trained on three discrete levels, the learned level embeddings
also support continuous interpolation: for $t\in[1,3]$ with $k=\lfloor t\rfloor$ and $\alpha = t-k$,
\begin{equation}
  \mathbf{e}(t) = (1-\alpha)\,\mathbf{e}_k + \alpha\,\mathbf{e}_{k+1},
  \label{eq:interp}
\end{equation}
which substitutes for $\mathbf{e}_\ell$ in Eq.~\eqref{eq:postfilter}, with gate bypassing at $t{=}1$.

\section{Experimental Settings}
\label{sec:data}

\noindent\textbf{Dataset.} We use MEAD~\cite{kaisiyuan2020mead}, a parallel
audio-visual corpus recorded across emotion categories and three intensity
levels. Disgust and contempt are excluded, as their facial action units and perception substantially overlap with other emotions \cite{DU_emotion}. Two speakers are also completely excluded due to unbalanced intensity coverage. The resulting corpus comprises 18{,}270 utterances, split speaker-disjointly into train/val/test of approximately 8:1:1. Each sample thus includes one face frame extracted from the source video, 22.050\,kHz audio, and a transcript. However, since
MEAD lacks ground-truth transcripts, we manually identify its underlying set of 141 canonical sentences, transcribe each utterance with Whisper-large~\cite{radford2023whisper}, and then correct transcriptions by matching against this set. 

\noindent\textbf{Training.} Starting from the LibriTTS pretrained backbone TTS model $G$,
we further train the style-conditioning branches for 30k steps with batch size 16 to learn emotion category and intensity from the face. This is then frozen, and the refinement module $R$ is trained separately for 15k steps. All training is conducted on a single NVIDIA GH200.

\begin{figure}[t]
  \centering
  \includegraphics[width=0.7\columnwidth]{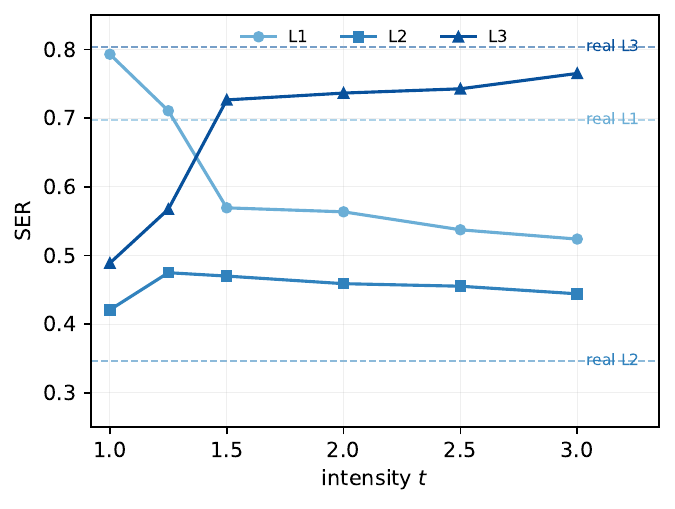}
  \caption{Continuous intensity adjustment from mild to strong emotion intensity. Dashed lines indicate real audio's recall.}
  \label{fig:dial}
\end{figure}

\noindent\textbf{Evaluation.}
We evaluate along four axes, combining objective and subjective metrics.
\textbf{(1) Intelligibility:} word error rate (WER; \%) between the input text and the
transcript of the synthesized speech, obtained via automatic speech
recognition with Whisper-large. \textbf{(2) Naturalness:} UTMOS \cite{utmos}, a
mean-opinion-score predictor for synthetic speech quality. \textbf{(3)
Emotion and Intensity Accuracy:} classification accuracy from a Wav2Vec2
speech-emotion recognizer fine-tuned on MEAD.
\textbf{(4) Human Evaluation:} overall TTS quality,
emotion-synthesis accuracy, and speaker transfer are assessed with 15 annotators who were recruited via Upwork.

\noindent\textbf{Baselines.} As this is the first attempt
at generating emotional speech at varying intensities directly from facial
input, no existing method targets the same task. We therefore adapt models from adjacent tasks for comparison. EmoSpeech~\cite{diatlova2023emospeech}
synthesizes emotive speech conditioned on an oracle ground-truth emotion
label at inference, serving as an upper-bound reference that bypasses visual
inference entirely. Furthermore, while FaceTTS~\cite{lee2023facetts} originally designed to predict voice identity from a face, it is repurposed here to predict emotion intensity from the face input.

\vspace{-5pt}
\section{Results}
\label{sec:res}
\vspace{-5pt}
Table~\ref{tab:combined_all} compares the proposed model with EmoSpeech~\cite{diatlova2023emospeech} and FaceTTS~\cite{lee2023facetts} using objective and subjective measures. Unlike EmoSpeech, which receives a ground-truth emotion label at synthesis time, our model infers both emotion category and intensity from a single facial image. Nevertheless, our model achieves higher unweighted intensity accuracy ($64.4$ vs.\ $39.1$), demonstrating improved intensity modeling without requiring emotion annotations at inference. Among the baseline systems, our model also achieves the lowest WER ($6.9\%$) and the highest UTMOS score ($3.14$).

Human evaluations corroborate these improvements. Our model obtains the highest naturalness MOS among the synthesis systems ($3.25$) and the highest emotion-synthesis accuracy at each of the three intensity levels. It also achieves a speaker-fidelity score of $4.13$. These results support the model's ability to convey the intended emotion while retaining the reference speaker's identity. 
The visualization in Figure~\ref{fig:clusters} provides complementary qualitative evidence, with generated utterances forming distinguishable clusters corresponding to the intended emotion categories.
Beyond synthesis at discrete intensity levels, the model also supports continuous intensity adjustment by interpolating the learned level embeddings without additional training. As seen in Figure~\ref{fig:dial}, increasing the intensity variable $t$ raises level-3 recall while reducing that of level-1.

\begin{table}[t]
\caption{Ablation of the intensity refinement module.}
\label{tab:intensity}
\centering
\footnotesize
\setlength{\tabcolsep}{2pt}
\begin{tabular*}{0.9\columnwidth}{@{\extracolsep{\fill}}lcccc@{}}
\toprule
System & L1 & L2 & L3 & Overall Acc. \\
\midrule
Real audio         & 69.7 & 34.6 & 80.3 & 61.5 \\
Backbone (no refinement)        & 79.3 & 42.1 & 48.9 & 56.7 \\
\textbf{+ refine., face-predicted} & 76.4 & 42.8 & 74.1 & 64.4 \\
\;+ refine., oracle label        & 79.3 & 45.9 & 76.5 & 67.2 \\
\bottomrule
\end{tabular*}
\end{table}

\begin{table}[t]
\caption{Fusion ablation. Speaker transfer is reported in percentages. The adopted methodology is indicated in grey.}
\label{tab:fusion}
\centering
\footnotesize
\setlength{\tabcolsep}{4pt}
\resizebox{0.9\columnwidth}{!}{%
\begin{tabular}{lcccc}
\toprule
Fusion & Emo. Acc.$\uparrow$ & Int. Acc$\uparrow$ & WER$\downarrow$ & Spk. Transfer$\uparrow$ \\
\midrule
cross-attn.       & \textbf{71.7} & 55.0 & 5.7 & 90.2 \\
frame cross-attn. & 61.6          & 48.3 & 8.1 & 50.2 \\
sum               & 71.2          & 55.2 & 5.8 & 90.8 \\
\rowcolor{gray!25}
concat$+$linear   & 70.3          & \textbf{56.7} & 6.6 & \textbf{92.1} \\
\bottomrule
\end{tabular}%
}
\end{table}

Furthermore, we quantify the effect of replacing ground-truth intensity labels with face-predicted levels during synthesis (Table~\ref{tab:intensity}). The resulting speech achieves an overall intensity accuracy of $64.4\%$, compared with $67.2\%$ under oracle conditioning, which is a small gap of $2.8$ percentage points. As such, to examine the contributions of individual components to the proposed model's performance, we conduct several ablation experiments. First, we evaluate how facial emotion and reference-audio vocal characteristics can effectively be fused into a shared conditioning vector (Table~\ref{tab:fusion}). Concatenation followed by linear projection achieves the highest intensity accuracy ($56.7$) before the intensity refinement module and speaker transfer ($92.1\%$), outperforming additive and attention-based fusion on these two objectives. This design preserves separate feature blocks before learning their joint contribution, consistent with the modalities' complementary roles: the face specifies affect, whereas the reference specifies speaker identity. As such, attention-based interactions may introduce unnecessary coupling between these roles.

\begin{table}[t]
\caption{Ablations with cross-speaker training.}
\label{tab:ablation2x2}
\centering
\footnotesize
\setlength{\tabcolsep}{4pt}
\resizebox{0.90\columnwidth}{!}{%
\begin{tabular}{lcccccc}
\toprule
Backbone & Refin. & Int. Acc$\uparrow$ & L3$\uparrow$ & SIM$\uparrow$ & WER$\downarrow$ & UTMOS$\uparrow$  \\
\midrule
paired        & --  & 57.2 & 43.7 & 77.9 & 5.2 & 3.17  \\
paired        & $+$ & 64.2 & 65.9 & 78.8 & 5.4 & 3.17  \\
cross-speaker & --  & 56.7 & 48.9 & 78.0 & 6.6 & 3.18  \\
\rowcolor{gray!25}
cross-speaker & $+$ & 64.4 & 74.1 & 78.5 & 6.9 & 3.14  \\
\bottomrule
\end{tabular}%
}
\end{table}

We further examine cross-speaker training, which pairs the reference audio with a facial image from a different speaker while matching emotion category and intensity. This strategy discourages reliance on facial identity for voice reconstruction, encouraging the model to derive affect from the face and timbre from the reference audio. As shown in Table~\ref{tab:ablation2x2}, cross-speaker training yields an increase in intensity accuracy over paired training and a larger improvement in level-3 intensity ($65.9\%$ to $74.1\%$).

\vspace{-5pt}
\section{Conclusion}
\vspace{-5pt}
In this paper, we have presented EmoFI-TTS, a framework for expressive speech synthesis that infers emotion category and intensity from a single facial image while using reference audio to specify speaker characteristics. Cross-speaker training encourages separation of affect and vocal identity, while intensity-conditioned refinement improves emotional expressiveness and supports continuous intensity adjustment without additional training. Evaluations on speaker-disjoint MEAD data demonstrate improved intensity accuracy, intelligibility, and naturalness over the evaluated baselines. Together, these findings establish facial affect as a promising conditioning signal for fine-grained emotional speech synthesis without emotion annotations at inference.

\vspace{-5pt}
\section{ACKNOWLEDGMENTS}
\label{sec:ack}
\vspace{-5pt}
This research was supported in part by the Natural Sciences and Engineering Research Council (NSERC) of Canada and in part by the AI2050 program at Schmidt Sciences. Finally, we are grateful for the support from IVADO and the Canada First Research Excellence Fund.

\section{Compliance with Ethical Standards}
This is a modeling study for which no ethical approval was required.

\bibliographystyle{IEEEbib}
\bibliography{strings,refs}

@inproceedings{popov2021gradtts,
      title={{Grad-TTS: A Diffusion Probabilistic Model for Text-to-Speech}}, 
      author={Vadim Popov and Ivan Vovk and Vladimir Gogoryan and Tasnima Sadekova and Mikhail Kudinov},
      year={2021},
      booktitle={Proceedings of the 38th International Conference on Machine Learning},
      url={https://arxiv.org/abs/2105.06337}, 
}

@ARTICLE{li2018deepface,
author={Li, Shan and Deng, Weihong},
journal={ IEEE Transactions on Affective Computing },
title={{Deep Facial Expression Recognition: A Survey}},
year={2022},
volume={13},
number={03},
ISSN={1949-3045},
pages={1195-1215},
doi={10.1109/TAFFC.2020.2981446},
url = {https://doi.ieeecomputersociety.org/10.1109/TAFFC.2020.2981446},
publisher={IEEE Computer Society},
address={Los Alamitos, CA, USA},
month=jul}

@INPROCEEDINGS{multispeaker2,
      title={{Enhancing Zero-Shot Multi-Speaker TTS with Negated Speaker Representations}}, 
      author={Yejin Jeon and Yunsu Kim and Gary Geunbae Lee},
      year={2024},
      booktitle={The Thirty-Eighth AAAI Conference on Artificial Intelligence (AAAI-24)}
}

@inproceedings{jeon-etal-2025-prompt,
    title = {{Prompt-Guided Selective Masking Loss for Context-Aware Emotive Text-to-Speech}},
    author = "Jeon, Yejin  and
      Kim, Youngjae  and
      Lee, Jihyun  and
      Lee, Gary",
    editor = "Chiruzzo, Luis  and
      Ritter, Alan  and
      Wang, Lu",
    booktitle = "Findings of the Association for Computational Linguistics: NAACL 2025",
    month = apr,
    year = "2025",
    address = "Albuquerque, New Mexico",
    publisher = "Association for Computational Linguistics",
    url = "https://aclanthology.org/2025.findings-naacl.38/",
    doi = "10.18653/v1/2025.findings-naacl.38",
    pages = "638--650",
    ISBN = "979-8-89176-195-7"
}

@inproceedings{jeon-etal-2025-progressive,
    title = {{Progressive Facial Granularity Aggregation with Bilateral Attribute-based Enhancement for Face-to-Speech Synthesis}},
    author = "Jeon, Yejin  and
      Kim, Youngjae  and
      Lee, Jihyun  and
      Kim, Hyounghun  and
      Lee, Gary",
    editor = "Christodoulopoulos, Christos  and
      Chakraborty, Tanmoy  and
      Rose, Carolyn  and
      Peng, Violet",
    booktitle = "Findings of the Association for Computational Linguistics: EMNLP 2025",
    month = nov,
    year = "2025",
    address = "Suzhou, China",
    publisher = "Association for Computational Linguistics",
    url = "https://aclanthology.org/2025.findings-emnlp.152/",
    doi = "10.18653/v1/2025.findings-emnlp.152",
    pages = "2799--2811",
    ISBN = "979-8-89176-335-7"
}

@inproceedings{kaisiyuan2020mead,
author = {Wang, Kaisiyuan and Wu, Qianyi and Song, Linsen and Yang, Zhuoqian and Wu, Wayne and Qian, Chen and He, Ran and Qiao, Yu and Loy, Chen Change},
title = {{MEAD: A Large-Scale Audio-Visual Dataset for Emotional Talking-Face Generation}},
year = {2020},
isbn = {978-3-030-58588-4},
booktitle = {ECCV 2020},
pages = {700–717},
numpages = {18}
}

@inproceedings{ye2025emotionalfaceto,
      title={{Emotional Face-to-Speech}}, 
      author={Jiaxin Ye and Boyuan Cao and Hongming Shan},
      year={2025},
      booktitle={ICML},
      url={https://arxiv.org/abs/2502.01046}, 
}

@INPROCEEDINGS{shi2025emotionAligned,
  author={Shi, Jiacheng and Du, Hongfei and He, Yangfan and Hong, Y. Alicia and Gao, Ye},
  booktitle={ICASSP}, 
  title={{Emotion-Aligned Generation in Diffusion Text to Speech Models Via Preference-Guided Optimization}}, 
  year={2026},
  pages={16492-16496},
  doi={10.1109/ICASSP55912.2026.11464274}}

@article{min2021metastylespeech,
  title={{Meta-StyleSpeech : Multi-Speaker Adaptive Text-to-Speech Generation}},
  author={Dong Min and Dong Bok Lee and Eunho Yang and Sung Ju Hwang},
  journal={ICML},
  year={2021},
  volume={abs/2106.03153},
  url={https://api.semanticscholar.org/CorpusID:235359041}
}

@inproceedings{desplanques2020ecapa,
   title={{ECAPA-TDNN: Emphasized Channel Attention, Propagation and Aggregation in TDNN Based Speaker Verification}},
   url={http://dx.doi.org/10.21437/Interspeech.2020-2650},
   DOI={10.21437/interspeech.2020-2650},
   booktitle={Interspeech 2020},
   publisher={ISCA},
   author={Desplanques, Brecht and Thienpondt, Jenthe and Demuynck, Kris},
   year={2020},
   month=Oct, pages={3830–3834},
   collection={interspeech_2020} }

@inproceedings{kong2020hifigan,
  author    = {Jungil Kong and Jaehyeon Kim and Jaekyoung Bae},
  title     = {{HiFi-GAN}: Generative adversarial networks for efficient and high fidelity speech synthesis},
  booktitle = {Proc. NeurIPS},
  year      = {2020}
}

@inproceedings{diatlova2023emospeech,
  author    = {Daria Diatlova and Vitaly Shutov},
  title     = {{EmoSpeech}: Guiding {FastSpeech2} Towards Emotional Text to Speech},
  booktitle = {Proc. 12th ISCA Speech Synthesis Workshop (SSW)},
  year      = {2023}
}

@inproceedings{lee2023facetts,
  author    = {Jiyoung Lee and Joon Son Chung and Soo-Whan Chung},
  title     = {Imaginary Voice: Face-styled Diffusion Model for Text-to-Speech},
  booktitle = {Proc. ICASSP},
  year      = {2023}
}

@inproceedings{belrose2023leace,
  author    = {Nora Belrose and David Schneider-Joseph and Shauli Ravfogel and Ryan Cotterell and Edward Raff and Stella Biderman},
  title     = {{LEACE}: Perfect linear concept erasure in closed form},
  booktitle = {Proc. NeurIPS},
  year      = {2023}
}

@inproceedings{baevski2020wav2vec,
  author    = {Alexei Baevski and Yuhao Zhou and Abdelrahman Mohamed and Michael Auli},
  title     = {wav2vec 2.0: A framework for self-supervised learning of speech representations},
  booktitle = {Proc. NeurIPS},
  year      = {2020}
}

@inproceedings{he2016resnet,
  author    = {Kaiming He and Xiangyu Zhang and Shaoqing Ren and Jian Sun},
  title     = {Deep Residual Learning for Image Recognition},
  booktitle = {Proc. CVPR},
  year      = {2016}
}

@article{ma2024emotion2vec,
  author  = {Ziyang Ma and Zhisheng Zheng and Jiaxin Ye and Jinchao Li and Zhifu Gao and Shiliang Zhang and Xie Chen},
  title   = {emotion2vec: Self-supervised pre-training for speech emotion representation},
  journal = {Proc. ACL Findings},
  year    = {2024}
}

@inproceedings{radford2023whisper,
  author    = {Alec Radford and Jong Wook Kim and Tao Xu and Greg Brockman and Christine McLeavey and Ilya Sutskever},
  title     = {Robust speech recognition via large-scale weak supervision},
  booktitle = {Proc. ICML},
  year      = {2023}
}

@INPROCEEDINGS {LSGAN_loss,
author = { Mao, Xudong and Li, Qing and Xie, Haoran and Lau, Raymond Y.K. and Wang, Zhen and Smolley, Stephen Paul },
booktitle = {ICCV},
title = {{Least Squares Generative Adversarial Networks}},
year = {2017},
volume = {},
pages = {2813-2821}}

@article{DU_emotion,
author = {Shichuan Du  and Yong Tao  and Aleix M. Martinez },
title = {Compound facial expressions of emotion},
journal = {Proceedings of the National Academy of Sciences},
volume = {111},
number = {15},
pages = {E1454-E1462},
year = {2014},
doi = {10.1073/pnas.1322355111},
URL = {https://www.pnas.org/doi/abs/10.1073/pnas.1322355111},
eprint = {https://www.pnas.org/doi/pdf/10.1073/pnas.1322355111}}

@inproceedings{utmos,
    author ={Takaaki Saeki and Detai Xin and Wataru Nakata and Tomoki Koriyama and Shinnosuke Takamichi and Hiroshi Saruwatari},
    title ={{UTMOS: UTokyo-SaruLab System for VoiceMOS Challenge 2022}} ,
    booktitle ={Interspeech},
    year = 2022
}

\end{document}